\documentclass[12pt]{article}
\usepackage{amssymb}
\usepackage{amsmath}
\usepackage{hyperref}
\begin{document}
\title{\bf  The Effect of Bulk Dimension in the presence of String Cloud on Viscosity Bound}

\author{Mehdi Sadeghi\thanks{Corresponding author: Email: mehdi.sadeghi@abru.ac.ir}  \hspace{2mm} \\
		 {\small {\em  Department of Physics, School of Sciences,}}\\
		 {\small {\em  Ayatollah Boroujerdi University, Boroujerd, Iran}}\\
    }

\date{\today}
\maketitle

\abstract{In this paper, the Einstein AdS black brane solution in the presence of a string cloud in the context of  d-dimensional massive gravity is introduced. The ratio of shear viscosity to entropy density for this solution violates the KSS bound by applying the Dirichlet boundary and regularity on the horizon conditions. Our result shows that this value is  independent of string cloud in any arbitrary dimensions.}\\



\section{Introduction} \label{intro}

Black holes are the solution of Einstein equation with an event horizon. In terms of the event horizon topology they are classified into three classes: Flat topology for black-branes, hyperbolic topology for topological black holes and spherical topology of the event horizon for black holes.
Black holes are interesting subject in high energy physics. Thermodynamics and phase transition of black holes are studied extensively in a few decades \cite{Witten:1998zw},\cite{Ranjbari:2019ktp},\cite{Ghanaatian:2019xhi}. Another interesting aspect of black holes arises in the context of AdS/CFT duality \cite{Maldacena97,Aharony99} where Einstein equation in d-dimensions can be reduced to hydrodynamics equation in $(d-1)$-dimensions \cite{Bhattacharyya07}. This map is called fluid/gravity duality \cite{Bhattacharyya07,Mateos07} in literature. The important implication of this duality is the study of  near-equilibrium property of strongly coupled systems. These fluids are strongly coupled and can not be described by perturbation methods, but  they can be described by holography, that is, by mapping them onto a weakly curved gravitational theory via fluid/gravity duality. In this duality,  Fluid is located at the boundary of AdS space-time. For an exact description of this fluid we should calculate the transport coefficients. The most important  coefficient is the ratio of shear viscosity to entropy density $\frac{\eta}{s}$. This value gets a lower bound ,$\frac{1}{4\pi}$, that for Einstein gravity is known as KSS bound \cite{Policastro2001,Policastro2002,Son2007,Kovtun2004,Kovtun2012}. This value is proportional to the inverse squared coupling of theory in boundary side. Membrane paradigm , pole method and Green-Kubo formula \cite{Policastro2001,Policastro2002} are the ways for calculation of $\frac{\eta}{s}$. Here, we calculate it by using Green-Kubo formula. Hydrodynamics equations for perfect fluid \cite{Kovtun2012} are the  conservation of $T^{\mu \nu}$ as follows,
\begin{align}
&\nabla_{\mu}T^{\mu \nu}=0, \\
&T^{\mu \nu}=\epsilon u^{\mu} u^{\nu} + p P^{\mu \nu}\nonumber,\\
&P^{\mu \nu}=\eta^{\mu \nu}+u^{\mu}u^{\nu}\nonumber,\\
&\eta^{\mu \nu}=(-1,1,1,1)\nonumber.
\end{align}
and equation of state $\epsilon=-p+Ts$ where $T$ is temperature and $s$ is entropy density. To illustrate the effects of dissipation we should add additional parts to the $T^{\mu \nu}$ as the following,
\begin{align}
& T^{\mu \nu } =(\epsilon +p)u^{\mu } u^{\nu } +pg^{\mu \nu } -\sigma ^{\mu \nu },\\
&\sigma ^{\mu \nu } = {P^{\mu \alpha } P^{\nu \beta } } [\eta(\nabla _{\alpha } u_{\beta } +\nabla _{\beta } u_{\alpha })+ (\zeta-\frac{2}{3}\eta) g_{\alpha \beta } \nabla .u].
\end{align}
\indent where $\eta$, $\zeta $, $\sigma ^{\mu \nu }$ and $P^{\mu \nu }$ are shear viscosity, bulk viscosity, shear tensor and projection operator, respectively \cite{Kovtun2012}. Green-Kubo formula can be derived by linear response theory as follows, 
\begin{equation}\label{Kubo}
\eta =\mathop{\lim }\limits_{\omega \to 0} \frac{1}{2\omega } \int dt\,  d\vec{x}\, e^{\imath\omega t} \left\langle [T^{x_1}_{x_2} (x),T^{x_1}_{x_2} (0)]\right\rangle =-\mathop{\lim }\limits_{\omega \, \to \, 0} \frac{1}{\omega } \Im G^{x_1\,x_1}_{x_2\, \, x_2} (\omega ,\vec{0}).
\end{equation}
The shear viscosity, $\eta$, is straightforwardly obtained from the Kubo formula and the entropy density, $s$, is obtained from the Bekenstein-Hawking formula.\\
 The basic elements of the nature  are considered  one-dimensional objects in string theory. The study of gravitational effects by string cloud is valuable. String cloud as a gravitational source studied for the first time in Einstein gravity \cite{Letelier:1979ej} and in Lovelock gravity  \cite{Herscovich:2010vr,Toledo} as a generalization of Einstein gravity. In \cite{Sadeghi:2019muh} we studied string cloud as a gravitational source in massive gravity in 4-dimensions, but in this paper, we will consider it in d-dimensions.\\
Hierarchy problem and the brane-world gravity solutions \cite{Dvali2000,Dvali2000-} predict the existence of massive graviton. There are some problems such as the cosmological constant problem and the current acceleration that GR can not explain them. For these reasons GR must be modified. Massive gravity \cite{deRham:2010kj} is one of the generalization of Einstein gravity in which graviton is not massless and this theory enjoys of ghost-free.  Massive gravity could explain the current observations related to dark matter \cite{Babichev:2016bxi} and also the accelerating expansion of universe without requiring any dark energy component \cite{Akrami:2012vf,Akrami:2015qga}. C. Bachas and et al \cite{Bachas:2018zmb} show that massive gravity can be embedded in string theory. In this paper, we consider d-dimensional AdS solutions of black hole solution in massive gravity in the presence of sting cloud.\\
Massive gravity can help to shed light on the quantum gravity effects\cite{Vasiliev96}.
In this paper, our goal is further exploration of string cloud in the framework of
massive theories of gravity.\\
Here, we consider massive gravity with string cloud \cite{Letelier:1979ej,Herscovich:2010vr,Toledo,Richarte:2007bx,Yadav:2009zza,Ganguly:2014cqa,Bronnikov:2016dhz,Barbosa:2016lse,Ghosh:2014pga,Lee:2014dha,Mazharimousavi:2015sfo,Graca:2016cbd} in d-dimensions and introduce the black-brane solution. Finally, we study the effect of bulk dimension in the presence of a string cloud on the value of $\frac{\eta}{s}$ and suggest some comments about the fluid  dual to this gravity model. 
\section{The Einstein AdS Black Brane with String Cloud Background in Context of Massive Gravity }
 \label{sec2}
The action of this model is given by,
\begin{equation}\label{action}
I=\frac{1}{2}\int{d^dx\sqrt{-g}\Bigg[R-2\Lambda+m_g^2\sum_{i=1}^{d-2}{c_{i}\mathcal{U}_i(g,f)}\Bigg]}+T_p\int_{\Sigma}\sqrt{-\gamma} d\lambda^0 d\lambda^1,
\end{equation}
where $ R $ is the scalar curvature, $\Lambda=\frac{-d_1d_2}{2l^2}$ is cosmological constant, $d_i=d-i$, $l$  is the radius of AdS spacetime, $f$ is a fixed rank-2 symmetric tensor known as reference metric and $m$ is the mass parameter. $ c_i $'s are constants and $ \mathcal{U}_i $ are symmetric polynomials of the eigenvalues of the $ d\times d $ matrix $ \mathcal{K}^{\mu}_{\nu}=\sqrt{g^{\mu \alpha}f_{\alpha \nu}} $. We choose a spatial reference metric (in the basis $(t,r,x_1,...,x_{d_2})$){\cite{Dehghani:2019thq,Sadeghi:2015vaa}
\begin{equation}\label{ref-metric}
f_{\mu \nu} =(f_{sp})_{\mu \nu}= \text{diag}(0,0,c_0^2 h_{ij}).
\end{equation}
where $c_0>0$ and $h_{ij}=\frac{1}{l^2}\delta_{ij}$ 
\begin{align}\label{U} 
  & \mathcal{U}_1=[\mathcal{K}],\,\,\,\,\,\,\,\,\mathcal{U}_2=[\mathcal{K}]^2-[\mathcal{K}^2],\,\,\,\,\,\,\,\mathcal{U}_3=[\mathcal{K}]^3-3[\mathcal{K}][\mathcal{K}^2]+2[\mathcal{K}^3]\nonumber\\
  & \mathcal{U}_4=[\mathcal{K}]^4-6[\mathcal{K}^2][\mathcal{K}]^2+8[\mathcal{K}^3][\mathcal{K}]+3[\mathcal{K}^2]^2-6[\mathcal{K}^4]\nonumber \\
 &
 \mathcal{U}_5=[\mathcal{K}]^5-10[\mathcal{K}^2][\mathcal{K}]^3+20[\mathcal{K}^3][\mathcal{K}]^2-20[\mathcal{K}^2][\mathcal{K}]^3+15[\mathcal{K}][\mathcal{K}^2]^2-30[\mathcal{K}][\mathcal{K}]^4+24[\mathcal{K}^5]\nonumber \\
&...
\end{align}
The rectangular brackets denote traces: $[A]\equiv A^{\mu}\,_{\mu}$.
\begin{align}
&[A^p] := (A^{\mu}\,_{\mu})^p\nonumber\\
&[A]^p := A^{\lambda_1}\,_{\lambda_2}A^{\lambda_2}\,_{\lambda_3}...A^{\lambda_p}\,_{\lambda_1}\nonumber
\end{align}
  The last part in the action is called the Nambu-Goto action of a string and $({\lambda}^{0},{\lambda}^{1})$  is a parametrization of the worldsheet, $T_p$ is a positive quantity and is related to the tension of the string and $\gamma$ is the determinant of the induced metric \cite{Ghosh:2014pga,Lee:2014dha,Mazharimousavi:2015sfo,Graca:2016cbd}
\begin{equation}\label{gamma}
{\gamma _{ab}} = {g_{\mu \nu }}\frac{{\partial {x^\mu }}}{{\partial {\lambda ^a}}}\frac{{\partial {x^\nu }}}{{\partial {\lambda ^b}}}.
\end{equation}
by this definition, we can write the Lagrangian part of the string as,
\begin{equation}
{S_{NG}} = T_p\int\limits_\Sigma  {\sqrt { - \frac{1}{2}{\Sigma _{\mu \nu }}{\Sigma ^{\mu \nu }}} } d{\lambda ^0}d{\lambda ^1}
\end{equation}
where
\begin{equation}\label{sigma}
 {\Sigma ^{\mu \nu }} = {\epsilon ^{ab}}\frac{{\partial {x^\mu }}}{{\partial {\lambda ^a}}}\frac{{\partial {x^\nu }}}{{\partial {\lambda ^b}}}
\end{equation}
 is the space-time bi-vector and satisfied in the following identities,
\begin{align}
&\Sigma ^{\mu [\alpha } \Sigma ^{\beta \gamma]}=0\\
&\nabla_{\mu}\Sigma ^{\mu [\alpha } \Sigma ^{\beta \gamma]}=0\\
&\Sigma ^{\mu \sigma } \Sigma_{\sigma \tau}\Sigma^{\tau \nu}=\gamma \Sigma^{\nu \mu}.
\end{align}
Where the square brackets refer to antisymmetrization in the enclosed indices.  $\epsilon ^{ab}$ is two-dimensional Levi-Civita tensor,  $\epsilon ^{01}=-\epsilon ^{10}=1$ and $\epsilon ^{00}=\epsilon ^{11}=0$.\\
The energy-momentum tensor for a single of string cloud can be calculated by following formula,
\begin{equation}\label{T_s} 
T^{\mu \nu }\text{(string)}= -2{{\partial \cal{L}} \over {\partial {g_{\mu \nu }}}}= -\,\, T_p\frac{{{\Sigma ^{\mu \sigma }}\Sigma _\sigma ^\nu }}{{\sqrt { - \gamma } }}.
\end{equation}
We consider the energy-momentum tensor for string cloud as follows,
\begin{equation}\label{T} 
T^{\mu \nu }\text{(cloud)}= \rho\,\, \frac{{{\Sigma ^{\mu \sigma }}\Sigma _\sigma ^\nu }}{{\sqrt { - \gamma } }}.
\end{equation}
where $\rho$ is the density of the string cloud and  conservation of the energy-momentum tensor ${\nabla _\nu }{T^{\mu \nu }} = 0$  results in \cite{Graca:2016cbd},
 \begin{equation}\label{TS2}
 \nabla_{\mu}(\rho \Sigma ^{\mu \sigma})\frac{\Sigma_{\sigma}\,\,^{\nu}}{(-\gamma)^{1/2}}+\rho \Sigma^{\mu \sigma} \nabla_{\mu}\big(\frac{\Sigma_{\sigma}\,\,^{\nu}}{(-\gamma)^{1/2}}\big) = 0.
 \end{equation}
By applying the same procedure of \cite{Sadeghi:2019muh} we have,
 \begin{equation}\label{T1}
 {\partial _\mu }(\sqrt { - g} \rho {\Sigma ^{\mu \sigma }}) = 0.
 \end{equation}
 In order to derive a d-dimensional AdS black hole solution with string cloud in massive theory of gravity, we consider the following static metric as an ansatz,
 \begin{equation}\label{metric}
 ds^{2} =-f(r)dt^{2} +\frac{dr^{2}}{f(r)} +r^2\sum_{i=1}^{d_2} h_{ij}dx^idx^j,
 \end{equation}
$h_{ij}dx^idx^j$ is the line element for an Einstein space
with constant curvature $d_2d_3 k$ with the values of  $k=1$  for spherical , $k=0$  for flat and $k=-1$  for  hyperbolic topology of the black hole horizon.\\
By using reference metric Eq.(\ref{ref-metric}) and background metric Eq.(\ref{metric}) , the values of $ \mathcal{U}_i $ are as below,
\begin{align}
	& \mathcal{U}_1=\frac{d_2c_{0}}{r}, \,\,\,  \,\,\, \mathcal{U}_2=\frac{d_2d_3c_0^2}{r^2},\,\,\,\,\mathcal{U}_3=\frac{d_2d_3d_4c_0^3}{r^3},\nonumber\\&\mathcal{U}_4=\frac{d_2d_3d_4d_5c_0^4}{r^4},...
\end{align}
$\mathcal{U}_i$'s can be written as $$\mathcal{U}_i=(\frac{c_{0}}{r})^i
\prod_{j=2}^{i+1}d_j,$$
in which $\displaystyle{\prod_{x}^{y}...=1}$ if $x>y$.\\
The field equations take the following forms by varying the action Eq. (\ref{action}) with respect to the spacetime metric,
  \begin{equation}\label{EoM}
  {G_{\mu \nu }} + \Lambda {g_{\mu \nu }} -{{m}^2}{\chi _{\mu \nu }} = T_{\mu \nu }\text{(string-cloud)}
  \end{equation}
where $G_{\mu \nu }= R_{\mu \nu}-\frac{1}{2}g_{\mu \nu}R$  is the Einstein tensor, $T _{\mu \nu }$ is the energy-momentum tensor of String Cloud and  ${\chi _{\mu \nu }}$  is massive term,
  \begin{align}\label{chi}
  \mathcal{\chi}_{\mu \nu} =\frac{c_1}{2}\bigg(\mathcal{U}_1 g_{\mu \nu }-\mathcal{K}_{\mu \nu}\bigg)+\frac{c_2}{2}\bigg(\mathcal{U}_2 g_{\mu \nu }-2\mathcal{U}_1\mathcal{K}_{\mu \nu}+2\mathcal{K}^2_{\mu \nu}\bigg)+\frac{c_3}{2}\bigg(\mathcal{U}_3 g_{\mu \nu }-3\mathcal{U}_2 \mathcal{K}_{\mu \nu}\nonumber\\+6\mathcal{U}_1 \mathcal{K}^2_{\mu \nu}-6\mathcal{K}^3_{\mu \nu}\bigg)+\frac{c_4}{2}\bigg(\mathcal{U}_4g_{\mu \nu }-4\mathcal{U}_3 \mathcal{K}_{\mu \nu}+12\mathcal{U}_2 \mathcal{K}^2_{\mu \nu}-24\mathcal{U}_1 \mathcal{K}^3_{\mu \nu}+24\mathcal{K}^4_{\mu \nu}\bigg)\nonumber\\+\frac{c_5}{2}\bigg(\mathcal{U}_5g_{\mu \nu }-5\mathcal{U}_4 \mathcal{K}_{\mu \nu}+20\mathcal{U}_3 \mathcal{K}^2_{\mu \nu}-60\mathcal{U}_2 \mathcal{K}^3_{\mu \nu}+120\mathcal{U}_1\mathcal{K}^4_{\mu \nu}-120\mathcal{K}^5_{\mu \nu}\bigg)+...
  \end{align}
Where $\mathcal{K}^n_{\mu \nu}=(\mathcal{K}_{\mu \nu})^n$.\\
The $rr$ component of field equation Eq.(\ref{EoM}) is,
\begin{equation}\label{eom}
d_2 d_3(k-f(r))-d_2 f'(r)r-2\Lambda r^2+m_g^2\bigg(\frac{c_0^ic_i}{r^{i-2}} \prod_{j=2}^{i+1} d_j\bigg)=2r^2 T_r ^r.
\end{equation}  
Where prime is differential with respect to radial coordinate. For the static and spherically symmetric string cloud the ansatz of the bivector $\Sigma^{\sigma \mu}$ is given by \cite{Herscovich:2010vr},
\begin{equation}\label{sigma1}
\Sigma^{\sigma \mu}=A(r)\bigg(\delta^{\sigma}_0 \delta^{\mu}_1-\delta^{\mu}_0 \delta^{\sigma}_1\bigg).
\end{equation}
By plugging Eq.(\ref{sigma1}) in Eq.(\ref{T}), the non-vanishing components of the energy-momentum tensor for the string cloud are as,
\begin{equation}
 T_t ^t=T_r ^r=-\rho|A(r)|.
\end{equation} 
By using Eq. (\ref{T1}) we get,
\begin{equation}
T^{\mu}_{ \nu}=-\frac{a}{r^{d_2}}\text{diag}[1,1,0,...,0],
\end{equation}
where $a$ is a positive constant.\\
By inserting the value of $T_r ^r$ in Eq. (\ref{eom}),
\begin{equation}
d_2 d_3(k-f(r))-d_2 r \frac{df(r)}{dr}-2\Lambda r^2+m_g^2\bigg(\sum_{i=1}^{d_2}\frac{c_0^ic_i}{r^{i-2}} \prod_{j=2}^{i+1} d_j\bigg)=\frac{-2a}{r^{d_4}},
\end{equation}
$f(r)$ is found as follows,
\begin{equation}\label{f1}
 f(r)=k-\frac{b}{r^{d_3}}+\frac{2a}{d_2r^{d_4}}-\frac{2\Lambda}{d_1 d_2} r^2+m_g^2\sum_{i=1}^{d_2} \bigg(\frac{c_0^ic_i}{d_2d_{i+1}r^{i-2}} \prod_{j=2}^{i+1} d_j\bigg).
\end{equation}
We know on event horizon $g^{rr}(r_0)=0$. If we consider this constraint on this solution, we can find the costant of $b$ as follows,
\begin{equation} 
b=r_0^{d_3}\bigg[k+\frac{2a}{d_2 r_0^{d_4}}-\frac{2\Lambda}{d_1 d_2} r_0^2+m_g^2\sum_{i=1}^{d_2} \bigg(\frac{c_0^ic_i}{d_2r_0^{i-2}d_{i+1}} \prod_{j=2}^{i+1} d_j\bigg)\bigg]. 
\end{equation}
By plugging the value of $b$ in $f(r)$ Eq.(\ref{f1}) we have,
 \begin{align}\label{f2}
&f(r)=\bigg[k\big(1-(\frac{r_0}{r})^{d_3}\big)-\frac{2a}{d_2r^{d_4}}(1-\frac{r_0}{r})-\frac{2r_0^2\Lambda}{d_1 d_2}\bigg((\frac{r}{r_0})^2-(\frac{r_0}{r})^{d_3}\bigg)\nonumber\\&+m_g^2\sum_{i=1}^{d_2} \bigg(\frac{c_0^ic_i}{d_{i+1}d_2}\big(\frac{1}{r^{i-2}}-\frac{1}{r_0^{i-2} }(\frac{r_0}{r })^{d_3}\big) \prod_{j=2}^{i+1} d_j\bigg)\bigg].
\end{align}
 We want to find the black-brane solution so we choose $k=0$. The emblackening factor follows as,
 \begin{align}\label{f}
 &f(r)=\bigg[-\frac{2a}{d_2r^{d_4}}(1-\frac{r_0}{r})-\frac{2r_0^2\Lambda}{d_1 d_2}\bigg((\frac{r}{r_0})^2-(\frac{r_0}{r})^{d_3}\bigg)\nonumber\\&+m_g^2\sum_{i=1}^{d_2} \bigg(\frac{c_0^ic_i}{d_{i+1}d_2}\big(\frac{1}{r^{i-2}}-\frac{1}{r_0^{i-2} }(\frac{r_0}{r })^{d_3}\big) \prod_{j=2}^{i+1} d_j\bigg)\bigg].
 \end{align}
We derive the entropy density for this solution by using of Hawking-Bekenstein relation,
 \begin{align}
&s=\frac{S}{V_{d_2}}=\frac{A}{4G V_{d_2}}=\frac{4 \pi}{V_{d_2}} \int{d^{d_2}x \sqrt{-g_{x_1x_1}...g_{x_{d_2}x_{d_2}}}}\bigg|_{r=r_0}\nonumber\\&=\frac{4 \pi}{V_{d_2}} \int{d^{d_2}x \sqrt{\chi}}=4 \pi \sqrt{\chi(r_0)}  =4\pi(\frac{r_0}{l})^{d_2},
\end{align}
where $V_{d_2}$ is the volume of the constant $t$ and $r$ hyper-surface with radius $r_{0}$ , $\chi(r_0)$ is the determinant of the spatial metric on the horizon and we used $\frac{1}{16\pi G} =1$ so $\frac{1}{4G} =4\pi$.
\section{Holographic Aspects of the Solution}
\label{sec3}
The important claim of AdS/CFT is the equivalence of partition function of gravity and gauge theories. It can be understood by GPK-Witten relation \cite{Aharony99} that stated:\\
 \begin{equation}
 \mathcal{Z_{\text{gauge}}}=\bigg<e^{i \int\phi^{(0)}\mathcal{O}}\bigg>=\mathcal{Z_{\text{string}}}=e^{i\bar{S}[\phi^{(0)}]}
 \end{equation}
where $\bar{S}$ is the on-shell action, $\phi$ is a field in the bulk theory and the $\phi^{(0)}$ is the value of bulk field in the boundary and considered as a external source both the boundary operator and bulk field. By using GPK-Witten prescription we calculate two-point Green’s function by differentiating the gravitational action with respect to the boundary values of fields then setting $\phi^{(0)}=0$,
 \begin{equation}
 <\mathcal{OO}>_s=\frac{\delta^2\bar{S}[\phi^{(0)}]}{\delta \phi^{(0)}\delta \phi^{(0)}}
 \end{equation}
2-point functions of energy-momentum tensor is related to shear viscosity from Kubo formula Eq.(\ref{Kubo}). Therefore, we perturb the bulk metric by $(\delta g){^{x_1}\,_{x_2}}=\phi(r)e^{-\imath\omega t}$.\\  
 We consider the metric and energy-momentum tensor that they are homogeneous and isotropic in the field theory directions as the following,  
 \begin{align}\label{Background}
 & ds^{2} =-g_{tt}(r)dt^{2} +g_{rr}(r)dr^{2}+g_{x_1x_1} (r) \sum_{i=1}^{d_2} dx^i dx_i,\\
 & T_{\mu \nu}=diag \bigg(T_{tt}(r),T_{rr}(r),T_{x_1x_1}(r),...,T_{x_{d_2}x_{d_2}}(r)\bigg).
 \end{align}
We apply stationary perturbation $ds^2=ds^2_0+2\phi(r)e^{-\imath\omega t}dx_1dx_2$ on the bulk metric and the perturbation equation reduces to \cite{Hartnoll:2016tri},
\begin{equation}\label{mode}
\frac{1}{\sqrt{-g}}\partial_r\bigg(\sqrt{-g}g^{rr}\partial_r{\phi}\bigg)+[g^{tt}\omega^2-m(r)^2]\phi=0,
\end{equation}
\begin{equation}
m(r)^2=g^{x_1x_2}T_{x_1x_2}-\frac{\delta T_{x_1x_2}}{\delta g_{x_1x_2}}\nonumber.
\end{equation}
Shear viscosity is calculated via Eq.(\ref{Kubo}),\\
\begin{equation}
\eta=\mathop{\lim }\limits_{\omega \, \to \, 0} \frac{1}{\omega } \Im G^{R} (\omega ,k=0)=\frac{\sqrt{\chi(r_0)}}{16\pi G_N}\phi_0(r_0)^2=\frac{s}{4\pi} \phi_0(r_0)^2.
\end{equation}
 Then, we will have,
\begin{equation}\label{formula}
\frac{\eta}{s}=\frac{1}{4\pi} \phi_0(r_0)^2.
\end{equation}
where $\phi_0$ is the solution of perturbation equation  Eq.(\ref{mode}) at zero frequency ($\omega=0$). We demand two conditions for $\phi$: it should be regular at horizon and the value of $\phi$ near the boundary is 1.\\
By considering this perturbation $\delta g_{x_1x_2}=\frac{r^2}{l^2}\phi(r)e^{i\omega t}$  on the bulk metric Eq.(\ref{metric}), we have,
\begin{equation}
ds^2=-\frac{f_1(r)}{l^2}dt^2 + \frac{l^2}{f_1(r)}dr^2 + \frac{r^2}{l^2}(dx_1^2 + dx_2^2 + 2\phi(r)dx_1dx_2+dx_3^2+...+dx_{d_2}^2),
\end{equation}
\begin{align}
f_1(r)&=\frac{l^2}{r^{d_3}}\bigg[\frac{2a}{d_2}(r-r_0)-\frac{2\Lambda}{d_1d_2}(r^{d_1}-r_0^{d_1})+m_g^2\frac{c_0c_1}{d_2}(r^{d_2}-r_0^{d_2})+m_g^2 c_0^2c_2(r^{d_3}-r_0^{d_3})\nonumber\\&+m_g^2c_0^3c_3(r^{d_4}-r_0^{d_4})d_3+m_g^2c_0^4c_4(r^{d_5}-r_0^{d_5})d_3d_4\bigg]=l^2 f(r).
\end{align}
By plugging this metric on the action Eq.(\ref{action}) and keeping up to second order of $\phi$ \cite{Policastro2001,Policastro2002,Son2007,Kovtun2004,Kovtun2012}, we have:
\begin{equation}\label{perturbed}
S_2=\frac{-1}{2} \int d^dx \Big(K_1 \phi'^2 -K_2 \phi^2\Big).
\end{equation}
we set $\omega=0$, where
\begin{align}
K_1&=\frac{r^{d_2}}{l^{d_2}}\frac{ f_1(r)}{l^{2}}=\frac{r^{d_2}}{l^{d}}f_1(r)=\frac{r}{l^{d_2}}\bigg[\frac{2a}{d_2}(r-r_0)-\frac{2\Lambda }{d_1d_2}(r^{d_1}-r_0^{d_1})\nonumber\\&+m_g^2 c_0^2c_2(r^{d_3}-r_0^{d_3})+m_g^2c_0^3c_3(r^{d_4}-r_0^{d_4})d_3+m_g^2c_0^4c_4(r^{d_5}-r_0^{d_5})d_3d_4\bigg] ,\\
K_2&= \frac{m_g^2}{2l^{d_2}}\big(c_0c_1 r^{d_3}+d_4c_0^2c_2r^{d_4}+d_4 d_5 c_0^3 c_3r^{d_5}\big)
\end{align}
Equation of motion for $\phi$ is found by variation of Eq. (\ref{perturbed}) with respect to $\phi$ then the EoM is,
\begin{equation}\label{EoM2}
(K_1 \phi')' + K_2 \phi=0.
\end{equation}
We try to solve the perturbation equation Eq.(\ref{EoM2}) perturbatively in terms of $m^2$ and  $a$. For the leading order we choose $m=a=0$ then the perturbation equation reduces to,
\begin{equation}\label{EoM-zeromass1}
\Bigg(r \left(r^{d_1}-r_0^{d_1}\right) \phi_0'(r)\Bigg)'=0,\,\,\,d \neq 1
\end{equation}
The solution is found,
\begin{align}
\phi_0(r)= C_1+C_2 \left(d \log r-\log \left(r r_0^d-r_0 r^d\right)\right)
\end{align}
By imposing regularity on the horizon and a Dirichlet boundary condition at the AdS boundary ,$\phi(r \to \infty)=1$, as boundary conditions we can find the constants $C_1=1$ and $C_2=0$. So the value of $\frac{\eta}{s}$ from (\ref{formula}) is given,
\begin{equation}\label{ValuM,a=0}
\frac{\eta}{s}= \frac{1}{4\pi}\phi(r_0)^2 = \frac{1}{4\pi}
\end{equation}
At the first order of perturbation we consider the solution as 
$\phi=1+m_g^2\phi_1(r)+a \phi_2(r)$ and put this solution to Eq.(\ref{EoM2}) and expand EoM in terms of powers of $m^2$ and $a$, we have,
  \begin{equation}
  \frac{m_g^2}{2l^{d_2}}\big(c_0c_1 r^{d_3}+d_4c_0^2c_2r^{d_4}+d_4 d_5 c_0^3 c_3r^{d_5}\big)-\left(\frac{2 \Lambda  m_g^2 r \left(r^{d_1}-r_0^{d_1}\right) \phi_1
  	'(r)}{l^{d}d_2 d_1}\right)'=0,
   \end{equation} 
  
\begin{align}
a\Bigg(r\left( r^{d_1}- r_0^{d_1}\right) \phi_2''(r)+\left(d  r^{d_1}-r_0^{d_1}\right) \phi_2'(r)\Bigg)=0.
\end{align} 
The solutions are found,
  \begin{align}
\phi_1 (r) \to C_2 + \int^r  \frac{C_1}{ r_0^d U-r_0 U^d} \, dU-\int^r \frac{c_0 d_2 d_1 r_0 U^{d_4} \left(\frac{c_0 c_2 d_4
   U}{d_3}+\frac{c_1 U^2}{d_2}+c_0^2 c_3 d_5\right)}{4\Lambda \left(r_0^d U-r_0 U^d\right)} \, dU
 \end{align} 
 \begin{align}\label{Phi2}  
\phi_2(r)\to C_3+\frac{C_4 r_0^{-d} \left(d \log (r)-\log \left(r r_0^d-r_0 r^d\right)\right)}{d_1} .
 \end{align} 
 $\phi(r)$ is as follows,  
\begin{align}\label{Phi-}
\phi(r) =& \phi_0+a C_3+m_g^2 C_2 +\frac{ \left(a C_4+m_g^2C_1 \right) \left(d \log r-\log \left(r r_0^d-r_0 r^d\right)\right)}{r_0^d d_1}\nonumber\\&-m_g^2\int^r \frac{c_0 d_2 d_1 r_0 U^{d_4} \left(\frac{c_0 c_2 d_4
   U}{d_3}+\frac{c_1 U^2}{d_2}+c_0^2 c_3 d_5\right)}{4\Lambda \left(r_0^d U-r_0 U^d\right)} \, dU
  \end{align}
 where $\Phi_0=\phi_0+a C_3+m_g^2 C_2$.\\
Near horizon of  Eq. (\ref{Phi-}) gives us,
 \begin{align}
\phi(r) =& i \pi\Bigg[-\frac{a C_4+m_g^2C_1  }{d_1 r_0^d}+\frac{c_0 d_2 d_1 r_0^{d_3}m_g^2 \left(\frac{c_0 c_2 d_4
   r_0}{d_3}+\frac{c_1 r_0^2}{d_2}+c_0^2 c_3 d_5\right)}{4 \Lambda}\Bigg]\log (r-r_0)+ ...
\end{align} 
 ... means finite terms. Now we apply the boundary conditions to find $C_1$ , $C_4$ and $\Phi_0$ constants. Regularity on horizon condition is deleted $\log(r-r_0)$, so we must vanish the coefficient, 
\begin{align} 
 C_1\to \frac{c_0 r_0^d d_2 d_1^2 r_0^{d_3} \left(\frac{c_0 c_2 d_4
    r_0}{d_3}+\frac{c_1 r_0^2}{d_2}+c_0^2 c_3 d_5\right)}{4 \Lambda}-\frac{a
 	C_4}{m_g^2},
\end{align} 
By substituting the above result of $C_1$ into Eq. (\ref{Phi-}) implies that,
\begin{align}\label{phi4}
\phi(r)& =\Phi_0+m_g^2 \Bigg(\frac{c_0 r_0^d d_2 d_1^2 r_0^{d_3} \left(\frac{c_0 c_2 d_4
    r_0}{d_3}+\frac{c_1 r_0^2}{d_2}+c_0^2 c_3 d_5\right)}{4 \Lambda}  \left(d \log r-\log (r r_0^d-r_0 r^d)\right)\nonumber\\&-\frac {c_ 0 d_ 2 d_ 1 r_ 0 } {4\Lambda}\int^r \frac{ U^{d_4} \left(\frac{c_0 c_2 d_4
       U}{d_3}+\frac{c_1 U^2}{d_2}+c_0^2 c_3 d_5\right)}{ \left(r_0^d U-r_0 U^d\right)} \, dU\Bigg)=\Phi_0-m_g^2B(r).
\end{align} 
By applying $\phi(r \to \infty)=1$ as a second boundary condition on Eq. (\ref{phi4}) 
The second boundary condition is at $\phi(r=\infty)=1$, which gives,
\begin{equation}
\Phi_0= \phi(r \to \infty)+m_g^2B(r \to \infty)=1+m_g^2B(r \to \infty)
\end{equation}
So $\phi(r)$ in $r_0<r<\infty$ is found,
\begin{equation}
\phi(r)= 1+m_g^2B(r \to \infty)-m_g^2B(r).
\end{equation}
Thus for calculating $\frac{\eta}{s}$ up to the first order in terms of $a$ and $m^2$ we evaluate the value of $\phi(r)$ at $r=r_0$  and by using  Eq.(\ref{formula}) we will have,
 \begin{align}
 \frac{\eta}{s}&= \frac{1}{4\pi}\phi(r_0)^2 = \frac{1}{4\pi} \Bigg(1+2m_g^2\bigg(B(r \to \infty)-B(r_0)\bigg)+O(m_g^4) \Bigg).
 \end{align}
  It means the value of $\frac{\eta}{s}$ is independent of string cloud in arbitrary dimensions of bulk and it behaves like massive gravity. So KSS bound is violated for $c_i<0$ by applying regularity of the solution of perturbation equation at horizon and $\phi(r \to \infty)=1$ on the boundary of AdS for this solution. On the other hand, this bound is preseved for $c_i>0$. Violation of KSS bound in massive gravity is due to mass term in perturbation equation, whiles in higher derivative gravity it is due to causality \cite{Ref22,Ref23,Ref24}. This bound is also preserved in Horava-Lifshitz gravity\cite{Sadeghi:2019trg}. 
 \section{Conclusion}
\noindent We considered string cloud in d-dimensional massive gravity in AdS space and introduced the black brane solution of this model then studied the aspects of holographic dual of this solution by calculating the $\frac{\eta}{s}$. The physical interpretation of this value is the inverse squared coupling of field dual. Our outcome shows the fluid dual of this model is the same as massive gravity and it is independent of string cloud in any arbitrary dimensions.  It also shows that string cloud acts like a charge or matter comparison to \cite{Sadeghi:2018ylh,Wu:2018zoq,Sadeghi:2020lfe} results. If we consider string cloud as a matter, we will see that $\frac{\eta}{s}$ is independent of matter. There is a conjecture that states $\frac{\eta }{s}=\frac{1}{4\pi}$ for Einstein-Hilbert gravity, known as KSS bound \cite{Ref22} which it is violated for higher derivative gravity\cite{Sadeghi:2015vaa,Ref22,Ref23,Ref24,Parvizi:2017boc,Sadeghi:2018vrf} and massive gravity with $c_i<0$ but the reason of them is different. \\\\
\noindent {\large {\bf Acknowledgment} }  Special thanks to Shahrokh Parvizi, for motivating this problem and useful comments. 


\end{document}